\documentclass{ws-procs9x6}
\begin{document}

\title{Infrared behavior of the running coupling constant and
quarkonium spectrum}

\author{\underline{M. BALDICCHI} ~and ~G. M. PROSPERI}

\address{Dipartimento di Fisica dell'Universit\`{a} di Milano \\
and I.N.F.N., Sezione di Milano, Italy}

\maketitle

\abstracts{We study the effect of the infrared behavior
of the running coupling constant on the quark-antiquark
spectrum.}

The running coupling constant in QCD is given, up to one loop in
perturbation theory, by
\begin{equation}
  \alpha_{\rm s} ( Q^{2} ) = \frac{ 4 \pi }{ \beta_{0}
  \ln{ ( Q^{2} / \Lambda^{2} ) } }
\label{runcst}
\end{equation}
$ Q $ being the relevant energy scale,
$ \beta_{0} = 11 - \frac{2}{3}  n_{\rm f} $ and
$ n_{\rm f} $ the number of flavors
with masses smaller than $ Q $.
\begin{figure}[htbp!]
  \begin{center}
    \begin{picture}(350,160)
      \centerline{\epsfxsize=3.in\epsfbox{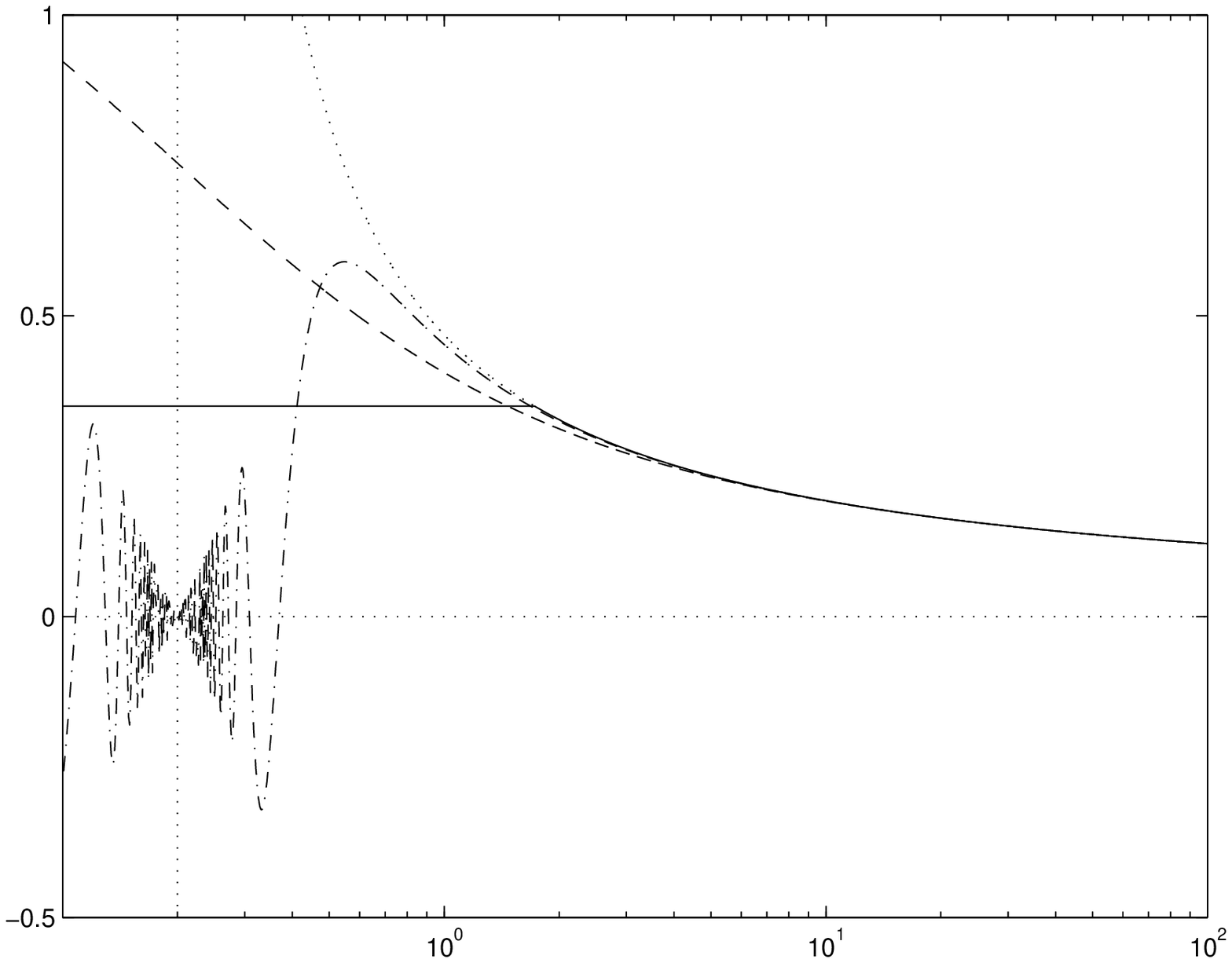}}
      \put(-280,150){ $ \alpha_{\rm s} $ }
      \put(-107,16){ $ Q $ (GeV) }
      \put(-247,-4){ $ \Lambda $ }
      \put(-285,96){ \small{0.35} }
    \end{picture}
  \end{center}
 \caption{Running coupling constant $ \alpha_{\rm s} (Q) $
 on logarithmic scale.
 Truncation prescription (full line),
 Shirkov-Solovtsov prescription (dashed line),
 Dokshitzer $ et \; al. $ prescription (dot-dashed line).}
 \label{figura}
\end{figure}
This expression has an unphysical singularity for
$ Q^{2} \to \Lambda^{2} $. Therefore it must be modified in
the infrared region.
\begin{table}[ph]
\tbl{$ q \bar{q} $ ($ q = u $ or $ d $).
$ n_{f} = 4, \; \Lambda = 0.2 \; {\rm GeV}. $
(a) truncation prescription,
$ m_{\rm u,d} = 0.01 \; {\rm GeV} $,
$ \alpha_{\rm s}(0) = 0.35 $,
$ \sigma = 0.2 \; {\rm GeV}^{2} $.
(b) Dokshitzer $ et \; al. \; \alpha_{\rm s}( Q^{2} ) $,
$ m_{\rm u,d} = 0.01 \; {\rm GeV} $,
$ \sigma = 0.18 \; {\rm GeV}^{2} $.
(c) Shirkov-Solovtsov $ \alpha_{\rm s}( Q^{2} ) $,
$ m_{\rm u,d} = 0.30 \; {\rm GeV} $,
$ \sigma = 0.18 \; {\rm GeV}^{2} $.
(d) Shirkov-Solovtsov $ \alpha_{\rm s}( Q^{2} ) $,
$ m_{\rm u,d}^{2} = 0.11 \,
k - 0.025 \, k^{2} + 0.265 \, k^{4} $,
$ \sigma = 0.18 \; {\rm GeV}^{2} $.\vspace*{1pt}}
{\footnotesize
\begin{tabular}{|c|c|c|c|c|c|c|}
\hline
States & {} & Experiment &  (a)  &  (b)  &  (c)  &  (d)  \\ [1ex]
  {}   & {} &   (MeV)    & (MeV) & (MeV) & (MeV) & (MeV) \\
\hline
$ 1 \, {^{1} S_{0}} $ &
$
\left\{
\begin{array}{c}
 \pi^{0} \\
 {} \\ [-1.5ex]
 \pi^{\pm}
\end{array}
\right.
$ & $
\left.
\begin{array}{c}
 134.9764  \pm 0.0006 \\
 {} \\ [-1.5ex]
 139.56995 \pm 0.00035
\end{array}
\right\}
$ &
479 & 575 & 473 & 124 \\
$ 1 \, {^{3} S_{1}} $ &
$ \rho(770) $ & $ 768.5 \pm 0.6 $ &
846 & 904 & 868 & 737 \\
$ ~~1 \, \Delta SS $ &  & 630 &
367 & 329 & 394 & 613 \\
$ 2 \, {^{1} S_{0}} $ &
$ \pi(1300) $ & $ ~1300 \pm 100 $ &
1326 & 1338 & 1326 & 1401 \\
$ 2 \, {^{3} S_{1}} $ &
$ \rho(1450) $ & $ 1465 \pm 25 $ &
1461 & 1459 & 1468 & 1508 \\
$ ~~2 \, \Delta SS $ &  & 165 &
135 & 121 & 142 & 107 \\
$ 3 \, {^{1} S_{0}} $ &
$ \pi(1800) $ & $ 1795 \pm 10 $ &
1815 & 1793 & 1806 & 1993 \\
$ 3 \, {^{3} S_{1}} $ &
$ \rho(2150) $ & $ 2149 \pm 17 $ &
1916 & 1889 & 1900 & 2063 \\
$ ~~3 \, \Delta SS $ &  & 354 &
101 & 96 & 94 & 70 \\
$
\begin{array}{c}
 1 \, {^{1} P_{1}} \\
 1 \, {^{3} P_{2}} \\
 1 \, {^{3} P_{1}} \\
 1 \, {^{3} P_{0}}
\end{array}
$
&
$
\begin{array}{c}
 b_{1}(1235) \\
 a_{2}(1320) \\
 a_{1}(1260) \\
 a_{0}(1450)
\end{array}
$
&
$
\begin{array}{c}
 1231 \pm 10 \\
 ~~~~ \left.
 \begin{array}{c}
  1318.1 \pm 0.7 \\
  ~~1230 \pm 40~ \\
  ~~1450 \pm 40~
 \end{array}
 \right\} 1303
\end{array}
$
&
1333 & 1365 & 1364 & 1319 \\
$
\begin{array}{c}
 1 \, {^{1} D_{2}} \\
 1 \, {^{3} D_{3}} \\
 1 \, {^{3} D_{2}} \\
 1 \, {^{3} D_{1}}
\end{array}
$
&
$
\begin{array}{c}
 \pi_{2}(1670)  \\
 \rho_{3}(1690) \\
      {}   \\
 \rho(1700)
\end{array}
$
&
$
\begin{array}{c}
 1670 \pm 20 \\
 \left.
 \begin{array}{c}
  1691.1 \pm 5 \\
     {}   \\
  1700 \pm 20
 \end{array}
 \right\}
\end{array}
$
&
1701 & 1715 & 1715 & 1741 \\
$
\begin{array}{c}
 1 \, {^{1} F_{3}} \\
 1 \, {^{3} F_{4}} \\
 1 \, {^{3} F_{3}} \\
 1 \, {^{3} F_{2}}
\end{array}
$
&
$
\begin{array}{c}
    {}  \\
 a_{4}(2040) \\
    X(2000) \\
    {}
\end{array}
$
&
$
\begin{array}{c}
    {} \\
 \left.
 \begin{array}{c}
  2037 \pm 26 \\
    {}  \\
    {}
 \end{array}
 \right\}
\end{array}
$
&
1990 & 1985 & 1979 & 2043 \\
$
\begin{array}{c}
 1 \, {^{1} G_{4}} \\
 1 \, {^{3} G_{5}} \\
 1 \, {^{3} G_{4}} \\
 1 \, {^{3} G_{3}}
\end{array}
$
&
$
\begin{array}{c}
      {} \\
 \rho_{5}(2350) \\
      {} \\
 \rho_{3}(2250)
\end{array}
$
&
$
\begin{array}{c}
  {} \\
 \left.
 \begin{array}{c}
  2330 \pm 35 \\
  {} \\
  {}
 \end{array}
 \right\}
\end{array}
$
&
2238 & 2214 & 2209 & 2319 \\
$
\begin{array}{c}
 1 \, {^{1} H_{5}} \\
 1 \, {^{3} H_{6}} \\
 1 \, {^{3} H_{5}} \\
 1 \, {^{3} H_{4}}
\end{array}
$
&
$
\begin{array}{c}
     {} \\
 a_{6}(2450) \\
     {} \\
     {}
\end{array}
$
&
$
\begin{array}{c}
     {} \\
 \left.
 \begin{array}{c}
  2450 \pm 130 \\
     {}  \\
     {}
 \end{array}
 \right\}
\end{array}
$
&
2460 & 2416 & 2415 & 2569 \\
\hline
\end{tabular}
\label{tabella} }
\vspace*{-13pt}
\end{table}
The most naive assumption consists in cutting the
curve (\ref{runcst}) at a certain maximum value
$ \alpha_{\rm s}(0) = \bar{\alpha}_{\rm s} $
to be treated as a mere phenomenological parameter
(truncation prescription).
Alternatively,
on the basis of general analyticity arguments,
Shirkov and Solovtsov \cite{shirkov} replace
(\ref{runcst}) with
\begin{equation}
  \alpha_{\rm s} ( Q^{2} ) = \frac{ 4 \pi }{
   \beta_{0} } \left(
  \frac{1}{ \ln{ ( Q^{2} / \Lambda^{2} ) } } +
  \frac{ \Lambda^{2} }{ \Lambda^{2} - Q^{2} } \right),
\label{runshk}
\end{equation}
since $ \alpha_{\rm s}(\Lambda^{2}) = 2 \pi / \beta_{0} $,
for $ Q^{2} \rightarrow \Lambda^{2} $ and
$ \alpha_{\rm s}(0) = 4 \pi / \beta_{0} $,
for $ Q^{2} \rightarrow 0 $ no singular point is left.
Finally, inspired also by phenomenological reasons,
Dokshitzer $ et \; al. $ \cite{lucenti} write
\begin{equation}
  \alpha_{\rm s} ( Q^{2} ) =
  \frac{ \sin ( \pi P ) }{ \pi P } \;
  \alpha^{0}_{\rm s} ( Q^{2} ),
\label{runlcnt}
\end{equation}
where $ \alpha^{0}_{\rm s} ( Q^{2} ) $ is the perturbative
running coupling constant as given by Eq. (\ref{runcst})
and
$ P = d/d ( \ln ( Q^{2} / \Lambda^{2} ) ) $
is a derivative acting on
$ \alpha^{0}_{\rm s} ( Q^{2} ) $.
The various curves are reported in Fig. \ref{figura}.
We solve the eigenvalue equation for a squared mass operator
$
M^{2} = M_{0}^{2} + U
$
obtained by reduction of a Bethe-Salpeter equation deduced
from first principle in QCD
under the only assumption that the
logarithm of the Wilson loop correlator $ W $ could be written as the
sum of its perturbative expression and an area term \cite{nora}
$
  i \ln W = i ( \ln W )_{\rm pert} + \sigma {\rm S}.
$
This formalism was successfully
applied in previous papers
\cite{miei}
to an overall evaluation of the spectrum in the light-light,
light-heavy and heavy-heavy sectors (the only
serious discrepancy with data
being for the light pseudoscalar meson masses).
In table \ref{tabella} we write the spectrum obtained for
the light-light systems.
With the truncation prescription (column (a)) and the
Dokshitzer $ et \; al. $
prescription (column (b)) we obtain good results
using a current quark mass $ m_{\rm u,d} = 0.01 \; {\rm GeV} $.
With the Shirkov-Solovtsov prescription (column (c)) we need to use a
higher constituent mass $ m_{\rm u,d} = 0.30 \; {\rm GeV} $.
A further improvement can be obtained if we use the
Shirkov-Solovtsov prescription with a running square mass
$ m_{\rm u,d}^{2} = 0.11 \,
k - 0.025 \, k^{2} + 0.265 \, k^{4} $ (column (d)),
in this case also the $ \pi $ mass is in the right range \cite{mio}.


\begin{thebibliography}{0}
\bibitem{shirkov}
  D. V. Shirkov, I. L. Solovtsov, {\it Phys. Rev. Lett.}
  {\bf 79}, 1209 (1997); \\
  {\it Theor. Math. Phys.} {\bf 120}, 1220 (1999); \\
  N. G.Stefanis, W. Schroers, H.-Ch. Kim, {\sl Phys. Lett.}
  {\bf B449}, 299 (1999); \\
  see also {\it Eur. Phys. J.} {\bf C18}, 137 (2000).
\bibitem{lucenti}
  Yu. L. Dokshitzer, in $ 29^{th} $ {\it International
  conference on High-Energy Physics} (ICHE 98), Vancouver, Canada,
  A. Astbury, D. Axen, J. Robinson Eds.
  World Scientific (1999), hep-ph/9812252;
 \\
  Yu. L. Dokshitzer, V.A. Khoze, S. I. Troyan,
  {\it Phys. Rev.} {\bf D53} 89 (1996).
\bibitem{nora}
  N. Brambilla, E. Montaldi, G.M. Prosperi, {\sl Phys. Rev.}
  {\bf D 54} (1996) 3506.
\bibitem{miei}
  M. Baldicchi, G.M. Prosperi, {\sl Phys. Rev.} {\bf D 62}
  (2000) 114024; \\
  {\sl Fizika} {\bf B 8} (1999) 2, 251;
  \\
  M. Baldicchi, in
  {\sl ``QCD: Perturbative or Nonperturbative ?''}
  Pag. 325,
  L. S. Ferreira, P. Nogueira, J. I.Silva-Marcos Eds.
  World Scientific (2001), hep-ph/9911268.
\bibitem{mio}
 M. Baldicchi, G. M. Prosperi, {\it Phys. Rev.} {\bf D66} 074008
 (2002).
\end{thebibliography}
\end{document}